\documentclass[preprint,aps]{revtex4}

\usepackage{epsfig,amsmath,graphics}

\newcommand{\be}{\begin{equation}}
\newcommand{\ee}{\end{equation}}
\newcommand{\bea}{\begin{eqnarray}}
\newcommand{\eea}{\end{eqnarray}}
\begin{document}
\title{Kohn-Sham equations for nanowires with direct current}
\author{D.S. Kosov}

\email{kosov@theochem.uni-frankfurt.de}

\affiliation{Institute of Physical and Theoretical Chemistry, \\ 
J. W. Goethe University, Marie-Curie-Str. 11, D-60439 Frankfurt, Germany}

\begin{abstract}
The paper describes the derivation of the Kohn-Sham equations for a nanowire
with direct current. A value of the electron current enters the problem 
as an input via a subsidiary condition imposed by pointwise Lagrange multiplier.
Using the constrained minimization of the Hohenberg-Kohn energy functional, 
we derive a set of self-consistent equations for current carrying orbitals 
of the molecular wire. 
\end{abstract}
\maketitle

\newpage

\section{Introduction}
The goal of building real electronic devices from individual 
molecules \cite{joachim2000} has spurred experimental 
and theoretical studies of molecular conductivities.
Measurements of current-voltage (I-V) characteristics 
have been reported for 
single molecules \cite{reed97,cui2001} but
mechanisms for electron transmissions through molecular interfaces 
remain largely unexplored and unexplained  \cite{nitzan2001,grossmann2002}. 
It appears that the miniaturization of microelectronic devices 
shrinks the size of the system to a level
when the simulations of transport characteristics 
of molecular electronic devices
can be performed by {\it ab initio} electronic structure 
methods (with single benzene-1,4-dithiolate molecule as the molecular 
wire, the contact wire-contact ``supermolecule'' consists of just 32 atoms). 
However, the main principle hurdle for the application of the 
first principle electronic structure methods is not the size of the system 
but conceptually unsolved problems of modeling of the electronic 
current and interpretation of the applied voltage for 
correlated electronic systems.
In  traditional methods, which are based on the Landauer theory 
\cite{landauer70},  the molecule and the edge part of the metal contacts
can be in principle treated {\it ab initio}, e.g. in the density functional 
theory (DFT).
But then to induce the electronic current it
 is assumed that the system 
is attached to two equilibrium electronic
reservoirs with the different chemical potentials \cite{datta95}.
Being computationally not very efficient, since it requires to 
perform the calculations over the whole coordinate space, these 
traditional approaches are also pivotally relied on the electronic 
reservoir concept and often
use  {\it ad hoc} reservoir-molecule interaction as an essential 
input parameter.

We have previously shown 
that the reservoir model can be conceptually circumvented
in electronic transport calculations if 
the electronic current 
is used as an input information instead the applied 
voltage bias \cite{kosov2001,kosov2002}. In our approach 
the electronic current enters  a quantum  variational
problem via the corresponding Lagrange multiplier. 
In this paper, we continue development of our method and discuss 
detailed derivation of the Kohn-Sham equations for a  
nanowire with steady current.  We begin by describing the variational 
principle for an inhomogeneous electron gas with steady current. We then derive
the Kohn-Sham equations with steady current.

\section{Variational principle}

We start our derivation by defining density and  current density of an 
inhomogeneous, interacting $N$-electron gas in an external 
scalar potential $v({\bf r})$ as (a.u. are used throughout the paper):
\begin{eqnarray}
\rho({\bf r}) &=& N  \int 
\Psi   ({\bf r},{\bf r}_2,\ldots;  {\bf r}_N)  
\Psi^* ({\bf r},{\bf r}_2,\ldots, {\bf r}_N) 
\prod_{i=2}^N d{\bf r}_i \; ,\\
\nonumber
 \\
{\bf j}({\bf r})&=& N 
\int  \mbox{Im}
\left\{  \Psi^* ({\bf r},{\bf r}_2,\ldots,  {\bf r}_N) 
\nabla \Psi({\bf r},{\bf r}_2,\ldots, {\bf r}_N )
\right\}
 \prod_{i=2}^N d{\bf r}_i \; , 
 \label{j}
\end{eqnarray}
where $\Psi$ is the many-electron wave-function. For simplicity of notation we do not write
explicitly the spin indeces but our derivation can be readily extended to
spin-polarized systems.

Next, we introduce a subsidiary condition for the current density distribution by
specifying the quantum wire
constraint on the current \cite{kosov2001,kosov2002}:
\begin{equation}
\int dy\, dz\, j_x ({\bf r}) = I(x) \;,
\label{constraint_wire}
\end{equation}
where $I(x)$  is a given input value of the steady current through the electronic  device.
With this constraint the current is aligned along the x-axis and 
the net current flow across a cross section
$\int dy\, dz\, j_x({\bf r})$ is constrained, and this
quantity is available experimentally.

For an inhomogeneous, interacting electron gas in scalar external potential $v({\bf r})$, the
ground state energy is given by the Hohenberg-Kohn energy functional \cite{hohenberg64}:
\begin{equation}
E_{HK}[\rho]=T_o [\rho ]+ \int d{\bf r} v({\bf r}) \rho({\bf r})+
\frac{1}{2} \int \int d{\bf r} d{\bf r'} \frac{ \rho({\bf r}) \rho({\bf r}')}
{|{\bf r}- {\bf r}'|} +E_{xc}[\rho ] \; .
\label{hk}
\end{equation}
The first term, $T_0$, is the kinetic energy functional of noninteracting electrons with 
given density and current density distribution. The second and third terms, 
are the interaction energy with 
external potential and the electrostatic interactions. 
The last term, $E_{xc}[\rho ]$, is the exchange and correlation energy functional.
The current dependent correction to the standard exchange-correlation 
energy functional is  $O(j^2) $ \cite{vignale87}. 
A dependence of the exchange-correlation energy functional  upon ${\bf j}({\bf r})$ 
can be readily included in our approach, yielding
the  additional exchange-correlation vector potential \cite{vignale87}:
\begin{equation}
{\bf A}_{xc}[\rho, {\bf j}] ({\bf r})= \frac{\delta E_{xc}[\rho, {\bf j}]}{\delta {\bf j}({\bf r})} \, 
\end{equation}
which modifies the kinetic energy operator in the final Kohn-Sham equation
in the same way 
as an external vector potential does. 
The current dependent correction to the standard exchange-correlation 
energy functional is  $O(j^2) $ \cite{vignale87}. 
Molecular and atomic wires
can sustain a current of about $10^{-4} a.u.$ for typical experiments on molecular conduction \cite{reed97}.
Therefore the direct current-dependent corrections to the total electronic 
energy are only of the order of $10^{-8} a.u.$ and can be ignored in most molecular transport calculations.

Likewise our derivation of the Schr\"odinger equation for current 
carrying states \cite{kosov2002},
we use the constrained variational principle in our derivations.
The problem is to minimize Hohenberg-Kohn energy functional (\ref{hk}) subject to the 
constraint on the current density (\ref{constraint_wire}).
The constraint is explicitly achieved by introduction of an auxiliary functional
\begin{equation}
\Omega[ \Psi]= E_{HK}[\rho]-
\int dx  a(x) 
\left(\int dy\, dz\, j_x ({\bf r}) - I(x) \right)
\; .
\label{omega}
\end{equation}
The first term giving the total energy is standard in the variational formulation 
of the DFT. The second term with the  Lagrange multipliers 
$a(x)$ has been introduced to impose the subsidiary constraint for
the current density (\ref{constraint_wire}).

\section{Kohn-Sham equations with direct current}

Following the Kohn and Sham approach \cite{kohn65}, we introduce a reference 
fermion system with orthonormal single-particle orbitals $\psi_i({\bf r})$ and 
occupation numbers $n_i$ to reproduce 
the charge and current densities:
\begin{equation}
\rho( {\bf r })= \sum_i n_i \psi_i^*({\bf r }) \psi_i({\bf r }) \; ,
\end{equation}
\begin{equation}
{\bf j}({\bf r })=\frac{1}{2i} \sum_{i} n_i \left( \psi_i^* ({\bf r }) \nabla \psi_i ({\bf r }) 
- \psi_i ({\bf r }) \nabla \psi_i^*({\bf r }) \right)\; .
\label{current_ks}
\end{equation}

According to the variational principle and assuming a fixed set of the occupation
numbers $n_i$, 
we carry out the minimization of the Hohenberg-Kohn energy functional with respect 
to the single-electron orbitals $\psi_i({\bf r})$ subject to the constraint for the current 
density (\ref{constraint_wire}), i.e. the variation of auxiliary functional $\Omega[\Psi]$.
The minimization yields the following 
single-electron self-consistent equations:
\begin{equation}
\left( -\frac{1}{2}\nabla + v({\bf r })+ \int d{\bf r }\frac{\rho({\bf r })}{|{\bf r }-{\bf r }'|}+
v_{xc}[\rho]({\bf r })+
 \frac{1}{2i}\left[ \partial_x, a(x) \right ]_+  
\right) \psi_i({\bf r}) =E_i\, \psi_i({\bf r}) \; ,
\label{ks1}
\end{equation}
where the anti-commutator term,
\be
\left[ \partial_x, a(x) \right ]_+ =\partial_x ( a(x) ) +2  a(x) \partial_x
\ee
is the additional imaginary potential arising directly from the
constraint on electronic current (Appendix A). 
This equation is a particular case of the more general equation derived 
recently by Kosov and Greer \cite{kosov2001}. 
Relating a position gradient with the corresponding momentum operator 
($ \hat{p}_x=-i \partial_x $)
we can rewrite the imaginary potential as the momentum-dependent operator:
\be
\frac{1}{2i} \left[ \partial_x, a(x) \right]_+ = \frac{1}{2}
\left[  \hat{p}_x, a(x) \right]_+ \; .
\label{ap}
\ee
Since the operator (\ref{ap}) is hermitian the
 Konh-Sham equation (\ref{ks1}) is a hermitian eigenproblem.

We have not specified yet the occupation numbers $n_i$ which enter
the Kohn-Sham equations (\ref{ks1}) via the density and the current density. 
To fix the set of occupation numbers,
we need to know the number of electrons in the nanowire. Although we deal with 
the open quantum system where electrons can leave and enter the wire, we 
still can put a constraint on the total number of electrons because
of time and space independence of the electronic current  
(i.e. steady state direct current). 
Then, the continuity equation tells us that the 
total number of electrons does not depend upon time. 
If we assume 
 that the charge neutrality is maintained 
after the establishment of the current currying state, 
the average number of electrons in the nanowire $N$ is the same
 as the number of electrons in the zero current wire.
Therefore to obtain the occupation numbers 
we can minimize $\Omega$ (\ref{omega}) with respect to  $n_i$ 
subject to the particle conserving constraint imposed with the additional 
Lagrange multiplier $\mu$ (chemical potential):
\be 
\delta\left[ \Omega - \mu ( \sum_i n_i -N) \right] \ge 0 \; .
\label{variation2}
\ee

The direct differentiation of the Hohenberg-Kohn energy is straitforward 
and results  into the following expression (Janak theorem) \cite{janak78}:
\bea
&&\frac{\partial E_{HK}}{\partial n_i} =
\int d{\bf r} \psi_i^*({\bf r}) \left( -\frac{1}{2} \nabla +v({\bf r}) +
\int d{\bf r }\frac{\rho({\bf r })}{|{\bf r }-{\bf r }'|}+
v_{xc}[\rho]({\bf r }) \right)  \psi_i({\bf r})  + 
\nonumber
\\
&& \left[ \sum_k n_k \int d{\bf r} \frac{\partial \psi_k^*({\bf r}) }{\partial n_i}
\left( -\frac{1}{2} \nabla +v({\bf r}) +
\int d{\bf r } \frac{ \rho({\bf r }) }{ |{\bf r }-{\bf r }'| }+
v_{xc}[\rho]({\bf r }) \right)  \psi_k({\bf r}) 
+c.c. \right]
\nonumber
\eea
The details of the differentiation of the constrained functional
with respect to an occupation number $n_i$ are given in Appendix B. The
differentiation leads to the following expression for the derivative:
\bea
\frac{\partial \Lambda}{\partial n_i}&& = \frac{1}{2i}
\int d{\bf r} \psi_i^*({\bf r}) \left[ \partial_x, a(x) \right]_+  \psi_i({\bf r})  + 
\nonumber
\\ 
&& 
\left[
 \frac{1}{2i} \sum_k n_k \int d{\bf r} \frac{\partial \psi_k^*({\bf r})}{\partial n_i}
\left[\partial_x, a(x) \right]_+ \psi_k({\bf r})
+c.c. \right] \; .
\eea
Combining these two derivatives we arrive to the following expression for
the derivative of the auxiliary functional:
\bea
&& \frac{\partial \Omega}{\partial n_i}  
=
\int d{\bf r} \psi_i^*({\bf r}) \left( -\frac{1}{2} \nabla +v({\bf r}) +
\int d{\bf r }\frac{\rho({\bf r })}{|{\bf r }-{\bf r }'|}+
v_{xc}[\rho]({\bf r })+ \frac{1}{2i} \left[\partial_x, a(x) \right]_+ \right)  \psi_i({\bf r})  + 
\nonumber
\\
&& \left[ \sum_k n_k \int d{\bf r} \frac{\partial \psi_k^*({\bf r})}{\partial n_i}
\left( -\frac{1}{2} \nabla +v({\bf r}) +
\int d{\bf r }\frac{\rho({\bf r })}{|{\bf r }-{\bf r }'|}+
v_{xc}[\rho]({\bf r })+ \frac{1}{2i} \left[\partial_x, a(x) \right]_+ \right)  \psi_k({\bf r})\right.
\nonumber
\\ 
&& \left.+c.c.\right] \; .
\eea
Given that  $\psi_i$ satisfy eq.(\ref{ks1}) it is easy to see that:
\be
 \frac{\partial \Omega }{\partial n_i}  
=E_i \int d{\bf r} \psi_i^*({\bf r}) \psi_i({\bf r}) 
+ \sum_k n_k E_k  \frac{\partial}{\partial n_i} \int d{\bf r}  \psi_k^*({\bf r}) \psi_k({\bf r})
=E_i \; .
\label{janak_theorem}
\ee 

From eq.(\ref{variation2}) with the use of eq.(\ref{janak_theorem}) 
we arrive to the inequality
\be 
\sum_i(E_i -\mu) \delta n_i \ge 0
\label{aufbau}
\ee
which gives rise to the standard ``build up principle''(
orbitals with $E_i < \mu$ have $n_i =1$ and those with 
$E_i > \mu$ have  $n_i =0$) which is still
hold within our approach:
a maximum two electrons  are put into current carrying 
orbitals in the order of
increasing orbital energy. 
The approach can be readily extended to the finite temperature case 
replacing of the integer $n_i=0 \mbox{ or } 1$ occupation number by the Fermi-Dirac
expression. The formal aspects of the  DFT with direct current and non-zero temperature 
are discussed elsewhere. 

The  Kohn-Sham equations with current (\ref{ks1}) are not yet in a form 
allowing for a solution to be found as the Lagrange multiplier  
$a(x)$ is not known yet. The additional equation for the Lagrangian multiplier
$ a(x)$ can be obtained if we require that the Kohn-Sham orbitals 
from the eq.(\ref{ks1}) yield the required current $I(x)$:
\begin{equation}
\frac{1}{2i} \sum_{i} n_i \int dy dz \left( \psi_i^* ({\bf r }) \partial_x \psi_i ({\bf r }) 
- \psi_i ({\bf r }) \partial_x \psi_i^*({\bf r }) \right)= I(x)\; .
\label{continuity}
\end{equation}
The system of nonlinear equations (\ref{ks1}, \ref{continuity}) are to be 
solved simultaneously and being combined together composes a set of 
Kohn-Sham equations for nanowire with direct current.

\section{Self-consistent absorbing-emitting boundary conditions}

Multiplying the eq. (\ref{ks1}) from the left on the  $n_i \psi_i({\bf r})^*$, summing up
over the Kohn-Sham orbitals 
and subtracting from the obtained equation its own complex conjugate
yields the continuity equation:
\begin{equation}
\nabla \cdot {\bf j}({\bf r}) = \partial_x ( a(x) \rho({\bf r})) \;.
\label{lagrange1}
\end{equation}
After the integration of  eq.(\ref{lagrange1}) over the $\{y,z\}$
plane
and substitution of the constraint for the current density eq.(\ref{constraint_wire})
into the equation 
the following first order differential equation 
for the Lagrange multiplier $a(x)$ can be deduced:
\begin{equation}
\partial_x (a (x) \rho_{yz}(x)  - I(x) )=0\; ,
\label{lagrange2}
\end{equation}
where we have introduced the quantity
\begin{equation}
\rho_{yz}(x) = \int dy dz \rho ({\bf r})\; .
\end{equation}
The solution of the differential  equation (\ref{lagrange2}) depends upon the asymptotic
behavior of the specified function for the current  $I(x)$. It is computationally convenient to
assume the space-localized boundary conditions ($I(x), \rho_{yz}(x)$ 
$\rightarrow$ 0, when  $x \rightarrow \pm \infty$).
This space-localized boundary conditions yields  zero integration constant 
for the differential equation (\ref{lagrange2}) and we obtain the following expression for the 
Lagrange multiplier:
\begin{equation}
a(x) = - \frac{I_x  }{\rho_{yz}(x)} \; .
\label{lagrange3}
\end{equation}
With this choice of the Lagrange multiplier (\ref{lagrange3})
we complete the Kohn-Sham equations with the current $I$ in
the self-consistent and closed form:
\begin{equation}
\left( -\frac{1}{2}\nabla + v({\bf r })+ 
\int d{\bf r }\frac{\rho({\bf r })}{|{\bf r }-{\bf r }'|}+
v_{xc}[\rho]({\bf r })
  - \frac{1}{2i} \left[\frac{I(x)}{\rho_{yz}(x)}, 
\frac{\partial}{\partial x} \right]_+ \right) \psi_i({\bf r})= 
E_i \psi_i ({\bf r}) \; .
\label{ks2}
\end{equation}
Some aspects of eq.(\ref{ks2}) deserve special discussion.
Suppose that the constrained current starts at the left system boundary L
and completely absorbed at the right system boundary R:
\begin{equation}
I (x) = I \; \theta(x-L) \theta(R-x)  \;,
\label{constraint-step}
\end{equation}
where $I$  is the steady current through the system
and $\theta$ is the step function.
Then one of the terms in the anticommutator imaginary potential
becomes proportional to the derivative of the step function
and yields the singular imaginary $\delta$-function potential on
the system boundaries:
$$
\frac{I i}{2 \rho_{yz}(x)} (\delta(x-L) - \delta(x-R))\;.
$$
This absorbing-emitting imaginary boundary potential is  optimized 
due to its self-consistency. It depends upon the 
density and value of the input current and emits 
the current $I$ at the left boundary and completely absorbs
the same current at the right boundary.

\section{Conclusions}
In this paper, we have derived the set of Kohn-Sham equations for a  
nanowire with direct, steady current. Not restricted to the linear 
response, our approach uses a constrained minimization of the total energy with a subsidiary 
condition for the current density to formulate the Kohn-Sham equation with 
direct current. The subsidiary condition for the current density is maintained during the variation
by a pointwise Lagrange multiplier. Being entirely based on the DFT, our method incudes 
electronic correlations into the transport calculations at the same level as they are taken
into account into the corresponding exchange-correlation functionals.

\newpage

\newpage
\appendix

\section{Functional derivatives of the constraint functional with respect to orbitals}
In this Appendix we compute the functional derivative of the
constraint functional.
The constraint functional has the following form
\begin{equation}
\Lambda[\psi_i, \psi^*_i] = \int dx  a(x) 
\left(\int dy\, dz\, j_x ({\bf r}) - I_x  \right)
\;, 
\label{constraint-functional}
\end{equation}
with
\be
j_x ({\bf r})= \frac{1}{2i} \sum_i n_i 
\left( \psi^*_i({\bf r}) \partial_x \psi_i({\bf r}) -
\psi_i({\bf r}) \partial_x \psi^*_i({\bf r}) \right)
\ee
The direct variation of the constraint functional with respect to the
$\psi^* ({\bf r})$ yields:
\begin{eqnarray}
\frac{\delta \Lambda[\psi]}{ \delta \psi_i^* ({\bf r})} &=&
\frac{1}{2i} \int d{\bf r}' a(x') \frac{ \delta}{ \delta \psi^* ({\bf r})}
\sum_k n_k \left\{ \psi_k^*({\bf r}') \partial_x \psi_k({\bf r}') 
- \psi_k({\bf r}') \partial_x \psi_k^*({\bf r}') \right\}
\nonumber
\\
&=&\frac{1}{2i}   a(x) \partial_x \psi_i({\bf r}) -
\frac{1}{2i} \frac{ \delta}{ \delta \psi_i^* ({\bf r})} 
\int d{\bf r}' \; a(x') \sum_k n_k \psi_k({\bf r}') \nabla \psi_k^*({\bf r}')
\nonumber 
\\
&=&\frac{1}{2i}  a(x) \partial_x \psi_i({\bf r}) -
\frac{1}{2i} \frac{ \delta}{ \delta \psi_i^* ({\bf r})} 
\int d{\bf r}' \; \partial_x \left \{ \sum_k n_k a(x') 
\psi_k({\bf r}') \psi_k^*({\bf r}') \right\}
\nonumber 
\\
&+&\frac{1}{2i} \frac{ \delta}{ \delta \psi_i^*({\bf r})}
\int d{\bf r}' \; \sum_k n_k  \psi_k^* ({\bf r}') 
\partial_x \left\{ a(x') \psi_k({\bf r}')\right\}
\nonumber 
\\
&=&
\frac{1}{2i}  a(x)  \partial_x \psi_i({\bf r})
+\frac{1}{2i} \partial_x \{ a(x) \psi_i({\bf r}) \} =
\frac{1}{2i} \left[ a(x), \partial_x \right]_+ \psi_x({\bf r})
\nonumber
\end{eqnarray}

\section{Differentiation of the constrained functional
with respect to an occupation number $n_i$}

In this appendix the technical details of the direct differentiation of
the constraint functional (\ref{constraint-functional}) is presented.
 
\bea
&&
\frac{\partial}{\partial n_i} \Lambda[\psi_i, \psi^*_i] = 
\frac{\partial}{\partial n_i} \int dx  a(x) 
\int dy\, dz\, j_x ({\bf r}) = 
\nonumber
\\
&&
\int dx  a(x) j_{i x}(x) +
\frac{1}{2i} \int dx  a(x) \int dy\, dz\, \sum_k n_k \frac{\partial}{\partial n_i}
\left( \psi^*_k({\bf r}) \partial_x \psi_k({\bf r}) -
\psi_k({\bf r}) \partial_x \psi^*_k({\bf r}) \right)
\eea
Here we have introduced the new quantity which is the current through the orbital $i$ 
integrated over the plane $\{y,z\}$:
\be
j_{i x}(x)=\frac{1}{2i}\int dy\, dz\, \left( \psi^*_i({\bf r}) \partial_x \psi_i({\bf r}) -
\psi_i({\bf r}) \partial_x \psi^*_i({\bf r}) \right)
\nonumber 
\ee

Then
\bea
&&
\frac{\partial}{\partial n_i} \Lambda[\psi_i, \psi^*_i] =  \int dx  a(x) j_{i x}(x) +
\frac{1}{2i} \sum_k n_k  \int d{\bf r}  a(x) 
\left( \frac{\partial \psi^*_k({\bf r})}{\partial n_i} \partial_x \psi_k({\bf r}) -
\frac{\partial \psi_k({\bf r})}{\partial n_i} \partial_x \psi^*_k({\bf r}) \right)
+
\nonumber
\\
&&
\frac{1}{2i} \sum_k n_k  \int d{\bf r}  a(x) 
\left( \psi^*_k({\bf r}) \partial_x \frac{ \partial \psi_k({\bf r})}{\partial n_i} -
\psi_k({\bf r}) \partial_x \frac{\partial \psi^*_k({\bf r})}{\partial n_i} \right)
\nonumber
\\
&&
 = 
\int dx  a(x) j_{i x}(x) +
\frac{1}{2i} \sum_k n_k  \int d{\bf r}  a(x) 
\left( \frac{\partial \psi^*_k({\bf r})}{\partial n_i} \partial_x \psi_k({\bf r}) -
\frac{\partial \psi_k({\bf r})}{\partial n_i} \partial_x \psi^*_k({\bf r}) \right)
+
\nonumber
\\
&&
\frac{1}{2i} \sum_k n_k  \int d{\bf r}   
\left( \frac{\partial \psi^*_k({\bf r})}{\partial n_i} \partial_x ( a(x) \psi_k({\bf r})) -
\frac{\partial \psi_k({\bf r})}{\partial n_i} \partial_x (a(x) \psi^*_k({\bf r})) \right)=
\nonumber
\\
&&
\int dx  a(x) j_{i x}(x) + \frac{1}{2i}
\left[ \sum_k n_k  \int d{\bf r}
\frac{\partial \psi^*_k({\bf r})}{\partial n_i}
[ \partial_x, a(x)]_+ \psi_k({\bf r}) - c.c. \right]=
\nonumber
\\
&&
\frac{1}{2i} \int d{\bf r} a(x)  \left( \psi^*_i({\bf r}) \partial_x \psi_i({\bf r}) -
\psi_i({\bf r}) \partial_x \psi^*_i({\bf r}) \right)
 + \frac{1}{2i}
\left[ \sum_k n_k  \int d{\bf r}
\frac{\partial \psi^*_k({\bf r})}{\partial n_i}
[ \partial_x, a(x)]_+ \psi_k({\bf r}) - c.c. \right]=
\nonumber
\\
&&
\frac{1}{2i} \int d{\bf r} \psi^*_i({\bf r}) [ \partial_x, a(x)]_+\psi_i({\bf r})
+
 \frac{1}{2i}
\left[ \sum_k n_k  \int d{\bf r}
\frac{\partial \psi^*_k({\bf r})}{\partial n_i}
[ \partial_x, a(x)]_+ \psi_k({\bf r}) - c.c. \right]
\eea

\clearpage


\end{document}